\documentclass[twocolumn,aps,prl,reprint,groupedaddress]{revtex4}
\usepackage{graphicx}
\usepackage{color}

\newcommand{\be}{\begin{equation}}
\newcommand{\ee}{\end{equation}}

\newcommand{\bea}{\begin{eqnarray}}
\newcommand{\eea}{\end{eqnarray}}

\newcommand{\p}{\partial}

\newcommand{\la}{\langle}
\newcommand{\ra}{\rangle}

\newcommand{\lp}{\left(}
\newcommand{\rp}{\right)}

\renewcommand{\Im}{{\rm \, Im\,}}
\renewcommand{\vec}[1]{{\bf #1}}
\renewcommand{\hat}[1]{{\widehat #1}}

\newcommand{\mpar}[1]{\marginpar{\small \it #1}}
\newcommand{\addJS}[1]{\textcolor{red}{#1}}

\begin{document}
\title{Energy-driven Drag at Charge Neutrality in Graphene}
%\title{Energy Transfer and Drag at Neutrality in Graphene Heterostructures}

\author{Justin C. W. Song$^{1,2}$}
\author{Leonid S. Levitov$^1$}
%\email{levitov@mit.edu}
\affiliation{$^1$ Department of Physics, Massachusetts Institute of Technology, Cambridge, Massachusetts 02139, USA}
\affiliation{$^2$ School of Engineering and Applied Sciences, Harvard University, Cambridge, Massachusetts 02138, USA}


%\author{}
%\thanks{}
%\affiliation{}


% \date{\today}

\begin{abstract}
Coulomb coupling
% Electron-electron scattering 
between proximal layers in graphene heterostructures results in 
%momentum and 
efficient energy transfer between the layers. We predict that, in the presence of correlated density inhomogeneities in the layers, 
vertical energy transfer has a strong impact on lateral charge transport. In particular, for Coulomb drag 
%regime 
it dominates over the conventional momentum drag near zero doping. The dependence on  doping and temperature, which is different for the two drag mechanisms, can be used to separate these mechanisms in experiment.
% or the energy contribution to drag is distinct from that for the conventional momentum drag.
% characteristic dependence on  doping and temperature, which is different for the momentum and energy contributions to drag, can be used to separate these mechanisms in experiment. 
We predict distinct features such as a peak at zero doping and a multiple sign reversal, which provide diagnostics for this new drag mechanism.
\end{abstract} 

\maketitle

Recently developed vertical heterostructures\cite{britnell1} comprised of a few graphene layers separated by an atomically thin insulating layer afford new ways to probe the effects of electron interactions at the nanoscale. Typical layer separation in these structures (1-2 nm)
can be very small compared to the characteristic electron lengthscales such as the de Broglie wavelength and the screening length. 
This defines a new strong-coupling regime wherein the interlayer and intralayer interactions are almost equally strong.
Fast momentum transfer between electron subsystems in the two layers
and strong Coulomb drag have been predicted in this regime\cite{tse2007,sensarma2010,narozhny2007,peres2011,katsnelson2011,narozhny2}
 with characteristic dependence on doping,
temperature and layer separation
distinct from that in previously studied systems\cite{gramila}.
%double quantum well systems\cite{gramila}.
Recently, measurements of strong drag in graphene were reported\cite{tutuc,geim}.

\begin{figure}
\includegraphics[scale=0.2]{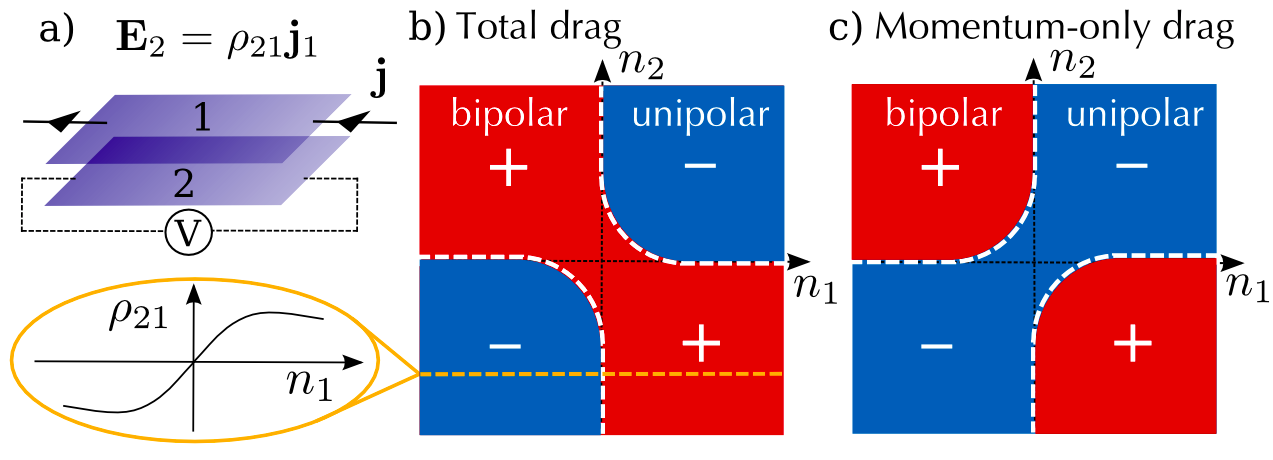}
\caption{Different mechanisms for Coulomb drag in graphene heterostructures.   Energy-driven drag 
%arises due to coupling of vertical energy trasfer and lateral charge transfer, 
dominates over  momentum drag near zero doping,
% dominates over momentum drag near the double charge neutrality (DCN), 
whereas momentum drag dominates at higher doping. The sign of the drag response depends on carrier polarity (a). For potential fluctuations of equal sign in the two layers, Eq.(\ref{eq:finitecorrelator}), the net drag (b) features a pair of nodal lines with an avoided crossing. The drag sign at zero doping is positive, as in the bipolar regime. This is distinct from momentum-only contribution to drag (c) smeared by correlated density fuctuations, $\delta\mu_1\approx\delta\mu_2$.
% Coulomb drag in graphene heterostructures (schematic). A current driven in layer 1 results in a voltage response in layer 2. (lower panel) Drag density dependence when one layer is fixed at high density (b) Schematic of drag resistivity dependence on carrier densities, $n_1$, $n_2$, with energy driven drag dominating at low doping. This results in an avoided crossing of the nodal lines.  (c) Schematic of carrier density dependent drag resistivity with density-smearing of momentum drag dominating at low dopings. This results in a distinctly different nodal structure as compared with panel (b).
%Momentum drag dominates at high doping, with the drag sign depending on carrier polarity, as illustrated by a slice taken 
% at density in one layer 
%away from CN. Energy-driven drag dominates at small doping, resulting in an avoided crossing of the nodal lines. % ``avoid each other'' near double CN where the energy drag dominates. 
%Slice of drag resistivity, $\rho_{21}$, taken at when the density in one layer is away from neutrality. (b)
% \addLL{New labels: b) Total drag (energy+momentum); c) Momentum-only contribution}
}
\label{fig1}
\vspace{-5mm}
\end{figure}

In this article we focus on another effect that becomes prominent in the strong coupling regime: vertical energy transfer mediated by interlayer electron-electron scattering. We predict that this process can give rise to lateral energy flow in the electron system, which under the conditions discussed below can directly impact electric transport. In particular, it leads to a characteristic contribution to the in-plane resistivity and dominates drag near charge neutrality (CN), see Fig.\ref{fig1} and Fig.\ref{fig3}.

Interlayer electron-electron scattering, which governs drag, is strongest at CN since the long-range Coulomb interactions become unscreened near CN. The scattering rate decreases with doping away from CN
as $\gamma \propto \nu^{-1}(\mu)$, where $\nu(\mu)$ is the massless Dirac density of states, $\mu\gg k_BT$ is the chemical potential, see Eq.(\ref{eq:gamma}). The $1/\nu$ scaling
is completely analogous to that found for intralayer scattering \cite{dassarma07,muller08}. 
Crucially, while both the energy and momentum transfer rates peak at CN, their impact on drag is markedly different. Since the sign of momentum drag depends on the polarity of charge carriers\cite{sivan}, momentum drag vanishes at CN \cite{tse2007,sensarma2010,narozhny2007,peres2011,katsnelson2011,narozhny2}. 
In contrast, energy-driven drag features a peak at CN, see Fig.\ref{fig3}.

Our drag mechanism arises due to coupling between vertical energy transfer and lateral charge and energy transport via spatial density inhomogeneity which is intrinsic to graphene.
Density inhomogeneity is known to be particularly strong at CN in the electron-hole puddle regime\cite{martin}, 
providing the dominant disorder potential in clean samples.
When a charge current is applied in layer 1, density inhomogeneity produces spatially varying heating/cooling [see Eq.(\ref{eq:jq})].
Strong thermal coupling between the electron systems in the two layers, mediated by the interlayer energy transfer, leads to a temperature pattern in layer 2 
that tracks that in layer 1, $\delta T_2(r)\approx \delta T_1(r)$.
Further, since
the disorder correlation length $\xi_{\rm dis}$ can reach 100 nm in G/BN heterostructures\cite{crommie1,crommie2}, exceeding the layer separation by orders of magnitude, 
the potential fluctuations are nearly identical in the two layers,
\be
\la \delta\mu_1(r) \delta\mu_2(r')\ra >0
\label{eq:finitecorrelator}
\ee
for $r\approx r'$.  As a result, the position-dependent thermopower induced by the gradient $\nabla\delta T_2(r)$ is correlated with the heating/cooling pattern in layer 1, giving rise to a nonzero ensemble-averaged drag voltage in layer 2. 

Our mechanism predicts a particular sign of the energy contribution to drag. 
% which 
%Also, interestingly, it has a sign
% which is identical to that of the bipolar regime away from CN. 
As a result, the density dependence for the net drag (energy and momentum combined) features a split-up pattern of nodal lines with an ``avoided crossing'' at zero doping,
%the double charge neutrality point (DCN), 
as illustrated in Fig. \ref{fig1} (b). The double sign change along the main diagonal $n_1=n_2$ and the peak at $n_{1,2}=0$
%DCN 
make the energy-driven drag easy to distinguish experimentally. 

As a parenthetical remark, the correlated density inhomogeneity, Eq.(\ref{eq:finitecorrelator}),  also affects the momentum drag, however its effect is opposite to that of the energy contribution (see Fig.\ref{fig1} b and c). If momentum drag were the dominant contribution near zero doping, the pattern of nodal lines would be such that the drag sign was constant along the main diagonal. This qualitative difference makes it easy to differentiate between the two cases.

% As a consequence of Eq. \ref{eq:finitecorrelator}, the case of energy-driven drag dominating at DCN produces nodal lines are separated by the downward sloping diagonal in Fig. \ref{fig1} (b). We remark that correlated density inhomogeneity between layers (Eq. \ref{eq:finitecorrelator}) can also smear momentum drag. However, in the case where momentum drag dominates at DCN, the nodal lines are separated by the upward sloping main diagonal as shown in Fig. \ref{fig1} (c). This qualitative difference makes it easy to differentiate between the two cases.

Below we develop a hydrodynamic framework to describe energy-driven drag.
The neutral modes (particle-hole excitations, or temperature imbalance) which mediate drag in our mechanism
 are of a long-range character, 
propagating over distances much larger than the inelastic mean free path $\ell=v/\gamma$. The length scales relevant for our drag mechanism are the electron-lattice cooling length and the inhomogeneity correlation length $\xi_T,\xi_{\rm dis}\gg\ell$ whichever is the smallest ($\xi_T$ can be as large as several microns, even at room temperature \cite{macdonald,wong,song}).
%  or  $\xi_{\rm dis}$, whichever is the smallest. 

% (we note similarity betwee our approach was developed in Ref.\cite{andreev})

\begin{figure}[t]
\includegraphics[scale = 0.26]{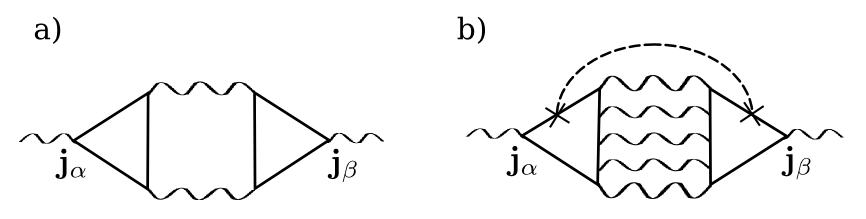}
\caption{Feynman diagrams for momentum (a)  and energy (b) contributions to drag. Wavy lines represent 
% interlayer 
interactions, dashed line represents disorder averaging. The ladder in (b) represents a long-wavelength charge-neutral mode.
%Diagram for momentum transfer contribution to drag, and (b) diagram for energy-driven contribution to drag 
% between layers
} 
\vspace{-5mm}
\label{fd}
\end{figure}

It is instructive to compare the Feynman diagrams describing the conventional momentum drag and our mechanism (see Fig.\ref{fd}). While the characteristic momenta are large ($\sim k_F$) for the former contribution (Fig.\ref{fd}a), 
the latter contribution (Fig.\ref{fd}b) includes ladder diagrams representing neutral modes propagating over distances of order $\xi_T$.
As a result, energy-driven drag is captured by a hydrodynamic framework which involves charge current $\vec j$ and heat current $\vec j_{\rm q}$, which in the ballistic transport regime are related by
\be\label{eq:jq}
\vec j_{\rm q} (\vec r)= Q (n) \vec j, \quad Q [n (\vec r) ] = \mathcal{S}[n (\vec r) ] T /e,
\ee
where $\mathcal{S}(n)$ is the entropy per carrier, $n(\vec r)$ is the density profile, and $e<0$ is carrier charge. 
In the ballistic regime, using electron temperature approximation, we find\cite{appendix}
% and negligible disorder-broadening of the Dirac point, we have
\be
Q =\frac{2\pi^2 k_B^2 T^2\mu}{3e (\mu^2 + \Delta^2(T))}.
 \label{eq:cleanq}
\ee
where $\Delta(T)$ accounts for the Dirac point broadening due to disorder  and thermal fluctuations.%The effects of Dirac point broadening by disorder, which are neglected in Eq.(\ref{eq:cleanq}), will be discussed below.

To illustrate the relation between energy and charge transport, we first analyze
% To illustrate the hydrodynamic approach, we first consider how our mechanism affects
%correlated heating/cooling and thermopower affect
in-plane resistivity in a single layer. According to Eq.(\ref{eq:jq}), spatial inhomogeneity leads to heating/cooling in the presence of uniform charge current (a la the Joule-Thomson process). The spatial temperature profile can be found from 
% an intra-layer heat equation 
$-\nabla \kappa \nabla \delta T + \lambda \delta T = - \nabla\cdot\vec j_{\rm q}$, where $\kappa$ is the thermal conductivity and $\lambda \delta T$ is the electron-lattice cooling power.
%, we obtain an inhomogeneous temperature profile, $\delta T (\vec r)$.  
A temperature gradient $\nabla\delta T$ drives thermopower, providing additional dissipation and thereby increasing resistivity. Onsager reciprocity combined with Eq.(\ref{eq:jq}) gives $\vec E(r) = - (Q[ n(\vec r)]/T)  \nabla \delta T$\cite{appendix}. Taking an ensemble average over small density fluctuations, $ \delta \mu \ll k_BT, \mu $, we find an increase in the in-plane resistivity, $\la \rho_{\alpha\beta}\ra  = \rho_{\alpha\beta}^0  + \Delta \rho_{\alpha\beta} $, [$\alpha (\beta) = x, y$], where
\be
\Delta \rho_{\alpha\beta} = \frac{1}{T} \sum_{|\vec q|\lesssim1/\ell}\frac{\la \delta Q (-\vec q) \delta Q(\vec q) \ra}{\kappa q^2 +  \lambda} q_{\alpha} q_{\beta}
.
\label{eq:inplane}
\ee
%where $\{\alpha, \beta\} = \{x, y\}$. 
%\addLL{What's the origin of the $\pi$ factor? Same in Eq.5 and Eq.14...}
%Evaluating the quantity $Q$ in the ballistic limit and negligible disorder-broadening of the Dirac point, we have
%\be
%Q =\frac{2\pi^2 k_B^2 T^2\mu}{3e (\mu^2 + (\pi^2/3) k_B^2 T^2)}.
% \label{eq:cleanq}
%\ee
% (the effects of broadening by disorder will be discussed below).
Since the derivative $\p Q/\p\mu$ peaks at $\mu=0$, this results in $\Delta\rho_{\alpha\beta}$ that peaks at CN. 
The temperature dependence estimated below is $\Delta \rho \propto T^2$, reminiscent of super-linear power laws for resistivity frequently observed at small doping\cite{chen}. A contribution of nonthermal modes to $\Delta\rho$ was analyzed in Ref.\cite{andreev}.
%\addLL{We estimate $\Delta\rho_{\alpha\beta}$ in the limit of small $\lambda$ (weak electron-lattice cooling), using $\kappa=(\mu^2+T^2)/T$ (see below). First, assuming $k_BT\ll\mu$, %we find
%\be
%\Delta \rho_{\alpha\beta}\sim \frac{T^6}{\mu^6\ell^2}\la\delta\mu_{q=0}^2\ra =\frac{T^{10}\ln^2 (T_0/T)}{\mu^8}
%\ee
%where we used an estimate $\ell=v/\gamma\sim v\mu/(T^2\ln (T_0/T))$, $T_0\sim\mu$ \addJS{and $\ell > \xi_{\rm dis}$}.
%To estimate $\Delta \rho$ near the Dirac point, we can set $T\sim\mu$, which gives $\Delta \rho_{\alpha\beta} \propto T^2$. 
%}
% \addJS{As we will show, in the disordered limit (DEFINE) and at not too low $T$ ($\ell < \xi_{\rm ds}$) the change in in-plane resistivity at small dopings goes as $\Delta \rho_{\alpha\beta} \propto T^2$.
%{\tiny{In the limit of small $\lambda$ (weak electron-lattice cooling), and using $Q$ in Eq.(\ref{eq:cleanq}) and $\kappa$ estimated later we find
%$\Delta \rho_{\alpha\beta} \propto T^2$. }}
% This is consistent with super-linear power laws for resistivity frequently observed at small doping, e.g. see Ref.\cite{chen}. }
% {\bf Please explain $T^$.}

% {\bf Trying: $\p Q/\p\mu\prop T^0$, $\kappa\rpto 1/T$ gives $\Delta \rho_{\alpha\beta} \propto T^{-2}$}

Generalizing this analysis to two layers coupled by vertical energy transfer and accounting for correlated density fluctuations, Eq.(\ref{eq:finitecorrelator}), we find an ensemble-averaged drag response $\vec E_2=\rho_{21}\vec j_1$, 
\be
\rho_{21} = \frac{1}{2T \tilde{\kappa}} \frac{\p Q}{\p\mu_1}\frac{\p Q}{\p\mu_2} \sum_\vec{q}\frac{\la \delta \mu_2 (-\vec q) \delta \mu_1 (\vec q) \ra}{1 +  \xi_c^2 \vec q^2}, 
% \quad \vec{E_2}= \rho_{21} \vec{j_1},
\label{eq:rho21}
\ee
where $\tilde\kappa = \kappa_1+\kappa_2$ is the net thermal conductivity of the two layers, 
$\mu$ is the chemical potential, 
%$\xi_c=\sqrt{ \kappa_1\kappa_2/[(\kappa_1+ \kappa_2)a] }$ 
$\xi_c\sim\ell$ is the interlayer cooling calculated below.
%length describing the interlayer thermal equilibration in  the heterostructure (see below).

Because the sign of the correlator in Eq.(\ref{eq:finitecorrelator}) is positive, energy-driven drag has the same sign as $\Delta \rho_{\alpha\beta}$ in Eq.(\ref{eq:inplane}), i.e. is {\it positive at zero doping}. This results in a double sign change along the main diagonal $n_1=n_2$,  as pictured in Fig. \ref{fig1} (b).
% This manifests in a total drag that is asymmetrical as pictured in Fig. \ref{fig1} (b). 
The density dependence for $\rho_{21}$ features a peak at zero doping  (see Fig.\ref{fig3}) which 
%is a hallmark 
is a hallmark of the energy-driven drag regime.

% As a parenthetical remark, 
We note that, if the sign of the correlator $\la \delta \mu_1 \delta \mu_2 \ra$ were negative, as expected for strain-induced charge puddles\cite{gibertini12}, our analysis carries through but predicts a negative drag at zero doping. Hence drag is a useful tool for probing the origin of inhomogeneity in graphene.

We begin by studying the energy transfer between the electronic systems in the two layers (Fig.\ref{fig1}(a)). This is described by the Hamiltonian
\be
\mathcal{H}= \sum_{i}\! \int\!\! d^2\vec r\psi_i^\dag (\vec r) \Big[ -i\hbar v \vec{\sigma} \cdot \vec{\nabla} + \delta \mu_i (\vec r) \Big]\psi_i(\vec r)  + \mathcal{H}_{\rm el-el} 
%\mathop{\sum_{\vec q, \vec k} }_{\vec{k'},i,j}V_{ij}(\vec q) \psi^\dag_{\vec k + \vec q, i} \psi^\dag_{\vec{k'} - \vec q,j} \psi_{\vec {k'},j} \psi_{\vec k, i} 
\label{eq:hamiltonian}
\ee
where $\{ i ,j\} = 1... 2N$ index layer, and spin/valley degrees of freedom, $\delta \mu (\vec r)$ describes the slowly varying disorder potential, $v$ is the Fermi velocity, and the  electron-electron interactions are $\mathcal{H}_{\rm el-el} = \sum_{\vec q, \vec k,\vec{k'},i,j}V_{ij}(\vec q) \psi^\dag_{\vec k + \vec q, i} \psi^\dag_{\vec{k'} - \vec q,j} \psi_{\vec {k'},j} \psi_{\vec k, i} $ with $V_{ij} (\vec q)$ the interaction. 

% In our subsequent
In our analysis, we ignore the correction due to finite layer separation $d$, approximating the interlayer interaction by the bare Coulomb interaction, $V_{ij} (\vec{q}) = 
%\frac{2\pi e^2}{\kappa |\vec q|}
V^0_\vec{q} e^{-d|\vec q| }\approx V^0_\vec{q} = 2\pi e^2/ \varepsilon |\vec q|$ with $\varepsilon$ the background dielectric constant. This approximation is valid when the lengthscale $d$ is small compared to the screening length and Fermi wavelength in the layers, which is the case for systems of current interest\cite{britnell1}.
% $q_0d\ll 1$, $q_0=-\Pi_1 (\vec q=0)- \Pi_2(\vec{q}=0)$ where $\Pi$ is the polarization function. 
%We will be interested in the regime when 
% Since typical values $d$ are small compared to the Fermi wavelength in the layers\cite{britnell1,britnell2}, the above approximation is adequate, 
The random-phase approximation then yields a screened interaction
% , $V (\vec q)$, reading  
$
 V_{ij}(\vec{q}) = V^0_{\vec q}/[1- V^0_{\vec q}( \Pi_1 (\vec q,\omega)+ \Pi_2(\vec{q,\omega}))]
$
for $i, j$ in different layers. 

We describe the energy distribution of carriers in each layer by a Fermi distribution at temperatures $T_{1,2}$.  Using Fermi's golden rule we can calculate the rate of energy exchange between the two layers (see Appendix). In the degenerate limit $\mu_1, \mu_2 \gg k_BT$, we obtain the energy loss power for layer 1 as 
\be
\mathcal{J}_{12} = \frac{6 \zeta(4)}{\hbar^3v^2} \frac{\nu_1\nu_2 k_B^4}{ (\nu_1 + \nu_2)^2}\Big( T_1^4 {\rm ln} \frac{T_0}{T_1} - T_2^4 {\rm ln} \frac{T_0}{T_2} \Big)
\label{eq:coolingpower}
\ee
where $\nu(\mu)$ is the total density of states in each layer, and $k_B T_0 = v (2\pi e^2/\varepsilon)(\nu_1+\nu_2)$.
%As expected, 
Notably, for equal densities $\mathcal{J}_{12}$ does not depend on the Fermi surface size. For equal densities and small temperature differences between the layers $T_1 \approx T_2$, we obtain the cooling rate
% can write the cooling rate as (scattering rate) as
\be
\gamma = \frac{1}{C_{\rm el}} \frac{d \mathcal{J}_{12}}{d T} =  \frac{9 \zeta(4) k_B^2 T^2}{ \pi \mu \hbar} {\rm ln} \frac{T_0}{T} %\frac{\big((T \, [{\rm K}]) / 300\big)^2 } {\mu [{\rm meV}] /100} {\rm ln}{ {\rm fs}^{-1}
\label{eq:gamma}
\ee
 where the heat capacity $C_{\rm el} = \pi^2/3 k_B^2 T \nu(\mu)$ and the density of states $\nu(\mu) = 2\mu/ (\pi \hbar^2 v^2)$ for the degenerate limit have been used. The rate $\gamma$ increases as $\mu$ goes towards neutrality, but is already quite large
%. Even 
for $\mu$ away from neutrality.
% fast cooling rates are obtained. In particular, for 
For typical values $\mu=100 \, {\rm meV}$, $T = 300\, {\rm K}$, the rate $\gamma$ is about $29 \, {\rm ps}^{-1}$,
%The non-degenerate limit produces a similar result (see Appendix). In particular at neutrality, and using the typical values for material constants displayed in the Appendix, cooling rates are
%\be
%\gamma  = \frac{T\, [{\rm K}] /300}{0.74} {\rm ps}^{-1}
%\label{eq:gamma}
%\ee
 orders of magnitude faster than the electron-lattice cooling rates\cite{macdonald,wong,song}.
% between the electronic system and lattice \cite{macdonald,wong,song}.
% Hence, there is little leakage of energy out of electron system into the lattice.
%  making hot carrier energy transport \cite{song2} in the heterostructure important.  

Vertical energy transfer couples heat transport in the two layers, so that the layer temperatures $T_1$, $T_2$ obey
\bea
&& -\nabla \kappa_1 \nabla \delta T_1+ a(\delta T_1- \delta T_2)+\lambda \delta T_1 = - \nabla\cdot \vec j_{q,1}
%\vec j_1 \cdot \nabla Q[n_1(\vec r)] 
%\label{eq:etransport1}
% \ee
%\be
\nonumber\\
&& -\nabla \kappa_2 \nabla \delta T_2+ a (\delta T_2-\delta T_1)+\lambda \delta T_2 = 0
\label{eq:etransport2}
\eea
where $a = d \mathcal{J}_{12}/ dT$ [see Eq.(\ref{eq:coolingpower})] and $\lambda$ describes electron-lattice cooling. We consider only a response linear in the applied current, $\vec j$, neglecting the quadratic joule heating term. 
Inverting the coupled linear equations, we find 
an increase in temperature in layer 2, $\delta T_2 (\vec r)$, that is driven by current in layer 1 as 
\be
\delta T_2 (\vec r)=  -  \frac{a}{\hat{L_1}\hat{L_2} - a^2}(\vec j_1 \cdot\nabla) Q[n_1(\vec r) ,T] ,
\label{eq:T2}
\ee
where $\vec j_{q,1}$ is the heat current, Eq.(\ref{eq:jq}), where $\hat{L}_i = -\nabla \kappa_ i\nabla + a +\lambda$. In what follows we suppress the $\lambda$ term since electron-lattice cooling is slow. Eq.(\ref{eq:T2}) then predicts a value for the interlayer cooling length
% in heterostructure, 
$\xi_c=\sqrt{ \kappa_1\kappa_2/[(\kappa_1+ \kappa_2)a] }$.
%length describing the interlayer thermal equilibration in  the heterostructure (see below).
The induced temperature profile, $\delta T_2 (\vec r)$, creates thermal gradients that can drive a local thermopower via $\vec E_2(r) = -(Q[ n_2(\vec r)]/T)  \nabla \delta T_2$.

Spatial fluctuations in  thermopower are governed by density fluctuations via Eq.(\ref{eq:T2}).
In particular, close to neutrality the local thermopower will exhibit regions of both positive and negative sign, leading to a spatial pattern of the drag resistivity. As discussed above, the correlations between $\delta\mu_1$ and $\delta\mu_2$, Eq.(\ref{eq:finitecorrelator}), lead to a nonzero ensemble-averaged drag resistivity. In the limit $ \delta \mu_{1,2} \ll k_BT, \mu_{1,2} $ we  write 
$Q_i(\vec r)  =\la Q_i(\vec r)\ra+\frac{\p Q}{\p\mu_i}\delta\mu_i(\vec r)$. Passing to Fourier harmonics via
$\la \delta\mu_1(\vec r)\delta\mu_2(\vec r')\ra  =  \sum_{\vec q}e^{i\vec q(\vec r-\vec r')}\la\delta\mu_1(-\vec q)\delta\mu_2(\vec q)\ra$, we obtain Eq.(\ref{eq:rho21}). 

\begin{figure}
%\flushleft
\includegraphics[scale=0.36]{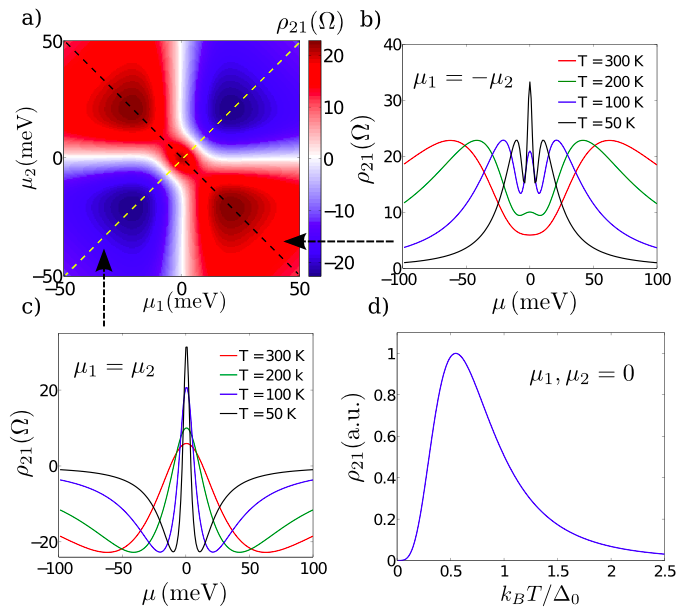}
\caption{ %Schematic of drag resistivity dependence on carrier densities $n_1$, $n_2$. (a) Nodal structure of drag with energy-driven drag dominating at low dopings (b) Nodal structure of drag with density-smearing of momentum drag dominating at low dopings.}
% Chemical potential dependence of combined 
 (a) Total drag resistivity $\rho^{\rm (tot)}_{21} = \rho^{\rm (m)}_{21} + \rho^{\rm (e)}_{21}$ {\it vs.} chemical potentials in the two layers, evaluated from Eq.(\ref{eq:momentum_drag}) and Eq.(\ref{eq:rho21}) 
% where $\rho^{\rm (m)}_{21} $ is the contribution from momentum transfer taken for 
at $T = 100\, {\rm K}$, producing a peak at $\mu_{1,2}=0$  (see text for parameter values used).
% The rest of the figures are for internal use. 
% We have used $\la \delta \mu^2 \ra \approx 25\, {\rm meV}^2$ and $\xi_{\rm dis} \approx 100\, {\rm nm}$ from \cite{crommie1,crommie2}.
(b,c) Slices $\mu_1 = \mu_2$ and $\mu_1 = -\mu_2$ at different temperatures.
% Chemical potential dependence of drag resistivity taken at slice denoted by black dotted line ($\mu_1 = -\mu_2$) in panel a at various temperatures showing three peaks at lower temperatures. (c)  Chemical potential dependence of drag resistivity taken at slice denoted by yellow dotted line ($\mu_1 = \mu_2$) in panel a at various temperatures showing a 
Note a three-peak structure in slice (b) and two sign changes close to CN in (c). (d) Temperature dependence of the peak at $\mu_{1,2}=0$.
%for disorder-smeared Dirac point obtained in the disordered limit, $\ell' < \ell$ (see text).
}
% of drag at DCN with disorder smearing of Dirac point and in the disordered limit, $\ell' < \ell$.
\vspace{-5mm}
\label{fig3}
\end{figure}

The fact that fluctuating local thermopower, exhibiting both positive and negative signs, does not average to zero % (in Eq. \ref{eq:rho21}) 
is surprising. 
% The correlated heating/cooling and thermopower conspire to give a non-vanishing $\rho_{21}$ since
This happens because the inhomogeneity in heat current and thermopower arise from the same source: electron-hole puddles. Energy-driven drag resembles mutual drag described by Laikhtman and Solomon \cite{laikhtman} in semiconducting heterostructures where doping at contacts produced a similar correlation between Peltier heating/cooling and thermopower. Energy-driven drag in Graphene differs from Ref.\cite{laikhtman} in that density inhomogeneity is intrinsic, occurs throughout the sample (not just at the contacts), and on a far smaller scale.

To see how energy-driven drag, Eq.(\ref{eq:rho21}), affects the total experimentally measured drag we need to account for momentum drag, $\rho^{\rm (m)}_{21}$.
We use a model that captures the main qualitative features of momentum drag:
% for $\rho_m$ 
\be\label{eq:momentum_drag}
\rho^{\rm (m)}_{21} = \tilde\rho^{\rm (m)}_{21} \frac{h}{e^2} (k_B T)^2 \frac{\mu_1}{(\mu_1^2 + \eta k_B^2T^2)}\frac{\mu_2}{(\mu_2^2 + \eta k_B^2 T^2)}
.
\ee
This expression, with the prefactor $\tilde \rho^{\rm (m)}_{21} = - 1.4\alpha^2/(2\pi \eta^2)$, the effective interaction strength $\alpha = 0.05$, and $\eta = 6.25$, was obtained by fitting the functional dependence derived in Ref. \cite{narozhny2} in the doping region $-10<\mu/k_{\rm B}T<10$. 
% corresponds to the functional dependence derived in Ref. \cite{narozhny2}. The parameter values   $\epsilon = 6.25$ and $\alpha = 0.05$ (effective interaction strength) are obtained
% a proportionality constant which we obtain from 
% by fitting the low doping region ($0<\mu<10k_{\rm B}T$) of the dependence in Ref. \cite{narozhny2}. 
% \addLL{The dependence 
% Here $\alpha$ is the interaction parameter and $\epsilon$ is the peak parameter. In the following we use values $\alpha = 0.05$ and the peak parameter $\epsilon = 6.25$ from Ref. \cite{narozhny2}.

%We use the values of interaction parameter $\alpha = 0.05$ to match typical experimentally measured drag resistivities reported in the literature \cite{tutuc,geim} and the peak parameter $\epsilon = 6.25$ obtained from the peak position in Ref. \cite{narozhny2}.

Combining this with $\rho^{\rm (e)}_{21}$ in Eq.(\ref{eq:rho21}), we obtain the total drag $\rho^{\rm (tot)}_{21} = \rho^{\rm (m)}_{21} + \rho^{\rm (e)}_{21}$
%$\rho_d = \rho_{21} + \rho_{m}$ 
plotted in Fig.\ref{fig3}. Here we have used $\kappa=(\mu^2 + \Delta^2(T))/ \hbar T$ \cite{lyakhov}%\cite{brooker,physicalkinetics}% \addLL{[{\bf need to justify}]}
, Eq.(\ref{eq:jq}) and assumed Gaussian correlations with average square density fluctuations $\la \delta \mu^2 \ra \approx 25\, {\rm meV}^2$ and $\xi_{\rm dis}=100\,{\rm nm}$ \cite{crommie1,crommie2}.
We note that the functional form of the correlator in Eq.(\ref{eq:finitecorrelator}) does not impact the qualitative behavior. %\addJS{$\kappa$ used above agrees with the Fermi Liquid thermal conductivity in the degenerate limit. In the non-degenerate limit, since $k_BT$ is the only relevant energy scale $\kappa \propto T$ which the above formula for $\kappa$ reproduces ($\epsilon \sim 1$).}
The obtained values of total drag 
% displayed in Fig. \ref{fig3}  generally 
are compatible with measured drag resistivities reported in Refs.\cite{tutuc,geim}. %\addJS{i'm not sure about the phrasing of this, my worry is that since narozhny momentum drag has peaks which don't change as a function of temperature, then it doesn't  ``agree" per se with Ref. \cite{geim}.} \addLL{we can change 'agree' to 'compatible' or kill the whole sentence}

%together with a Lorentzian correlator $\la \delta\mu_1(-\vec q)\delta\mu_2(\vec q)\ra = \la \delta \mu^2 \ra\xi_d^2/(1+ \xi_d^2 q^2)$. Here $\la \delta \mu^2 \ra \approx 25\, {\rm meV}^2$ \cite{crommie1,crommie2} is the average square potential fluctuation caused by density inhomogeneity in a single sheet. 
%We note parenthetically that 
 % The range of 

The density dependence of total drag plotted in Fig.\ref{fig3} (a) 
% serves as a clear way of identifying the two mechanisms for 
can be used to distinguish the two drag mechanisms in experiments. Namely, the peak at zero doping is due to energy-driven drag. On the slice $\mu_1=-\mu_2$ (black dashed line) this peak is surrounded by two peaks dominated by the momentum contribution [Fig.\ref{fig3}(b)]. On the slice $\mu_1=\mu_2$ (yellow dashed line) the two mechanisms produce contributions of opposite sign, resulting in a double sign change [Fig.\ref{fig3}(c)]. This provides a clear means of discerning the energy-driven regime.

% In particular, the peak at DCN shown in the density slice (black dashed line) is a consequence of thermal momentum, $Q$, peaking at CN (see Eq. \ref{eq:cleanq}). The resulting three peak pattern (Fig. \ref{fig3} (b)) is a hallmark of energy-driven drag. Furthermore, along the main diagonal (yellow dashed line), a double sign reversal in total drag (Fig. \ref{fig3} (c)) provides clear means of discerning the energy-driven regime.

%The above signatures for energy-driven drag in Fig. \ref{fig2} are distinct from density inhomogeneity smearing of the momentum drag. In particular, density smearing of density dependence in Eq.(\ref{eq:momentum_drag}) produces a pattern of nodal lines qualitatively different from of energy-driven drag (compare panels (a) and (b) in Fig. \ref{fig3})
%that instead link the unipolar regions (continuous blue region) and separate the bipolar regions (two distinct areas of red). 
%This provide a simple way to differentiate between our energy-driven effect and density-smearing of momentum drag.

%Close to neutrality, the temperature dependence of $\rho_{21}$ (Eq. \ref{eq:rho21}) and $\Delta \rho_{\alpha \beta}$  (Eq. \ref{eq:inplane}) are alike. To illustrate this, we consider first $\Delta \rho_{\alpha \beta}$. In the limit of weak $\lambda$ and 
The temperature dependence can be estimated as follows. At very low $T$ such that $\ell,\xi_c \gtrsim \xi_{\rm dis}$, the sum in Eq.(\ref{eq:rho21}) is cut at $1/\ell$, giving $\rho_{21} \propto T^{8}$.
%. Since $\ell, \xi_c \propto \sqrt {\Delta^(T)}/ T^2 $ (see Eq. \ref{eq:gamma}) we obtain $\Delta \rho_{\alpha\beta}, \rho_{21} \propto T^{8}$. 
At not too low $T$ such that $\xi_c,\ell \lesssim \xi_{\rm dis}$, 
the sum in Eq.(\ref{eq:rho21}) yields
%the $\sum_\vec{q}$ in Eq. \ref{eq:inplane} can be approximated to be constant by recalling 
$ \sum_{\vec q} \la\delta\mu_1(-\vec q)\delta\mu_2(\vec q)\ra = \la \delta\mu_1(\vec r)\delta\mu_2(\vec r')\ra _{\vec r = \vec r'}$.
% = \la \delta \mu^2 \ra$. 
This gives a non-monotonic $T$ dependence [see Fig. \ref{fig3} (d)]
% so that close to neutrality
\be
% \Delta \rho_{\alpha\beta} 
 \rho_{21} \propto \frac{T^4 }{\big(\Delta_0^2 + \eta (k_BT)^2\big)^3}
\la \delta\mu_1(\vec r)\delta\mu_2(\vec r')\ra _{\vec r = \vec r'}
.
\label{eq:tempdep}
\ee
%
% see Fig. \ref{fig3} (d).
% For $k_BT \ll \Delta_0$, this gives $\Delta \rho_{\alpha\beta} \propto T^4$. %consistent with super-linear resistivities frequently reported for small dopings \cite{chen}. 
% For $k_BT \gg \Delta_0$ this gives $\Delta \rho_{\alpha\beta} \propto T^{-2}$. 
A similar non-monotonic $T$ dependence arises for in-plane resistivity $\Delta \rho_{\alpha\beta}$.
% is also reproduced in drag resistivity at DNP as shown in Fig. \ref{fig3} (d). 
Interestingly, a peak in drag resistance at CN with non-monotonic temperature dependence was recently reported in Ref.\cite{geim}. 
%is a consequence of thermal and disorder broadening of the Dirac point and can be used as an additional signature of energy-driven drag.%where the condition is $\xi_c < \xi_d$. 
 % For the lowest temperatures, where $\ell,\xi_c > \xi_d$, both the sum in $\vec q$ in Eq. \ref{eq:inplane}  and \ref{eq:rho21} are cut by $1/\ell$ ($1/\xi_c$). Since $\ell, \xi_c \propto \sqrt {\Delta^(T)}/ T^2 $ (see Eq. \ref{eq:gamma}) we obtain $\Delta \rho_{\alpha\beta}, \rho_{21} \propto T^{8}$. 

%\addLL{We estimate $\Delta\rho_{\alpha\beta}$ in the limit of small $\lambda$ (weak electron-lattice cooling), using $\kappa=(\mu^2+T^2)/T$ (see below). First, assuming $k_BT\ll\mu$, we find
%\be
%\Delta \rho_{\alpha\beta}\sim \frac{T^6}{\mu^6\ell^2}\la\delta\mu_{q=0}^2\ra =\frac{T^{10}\ln^2 (T_0/T)}{\mu^8}
%\ee
%where we used an estimate $\ell=v/\gamma\sim v\mu/(T^2\ln (T_0/T))$, $T_0\sim\mu$ \addJS{and $\ell > \xi_{\rm dis}$}.
%To estimate $\Delta \rho$ near the Dirac point, we can set $T\sim\mu$, which gives $\Delta \rho_{\alpha\beta} \propto T^2$. 
%}

The above analysis can be easily extended to describe the diffusive limit where the elastic mean free path  is shorter than the inelastic mean  free path, $\ell'<\ell$. Our hydrodynamic approach remains valid in this regime, with the quantity $Q = sT$ where $s$ is the Seebeck coefficient. 
The energy-driven drag is still given by Eq.(\ref{eq:rho21}), 
% but with $Q = sT$ where the Seebeck coefficient, 
with $s$ and 
% thermal conductivity, 
$\kappa$ described by the Mott and Wiedemann-Franz relations:
% respectively:
\be\label{eq:Mott_formula}
s = \frac{\pi^2}{3e} k_B^2 T \frac{\partial {\rm ln} \sigma}{\partial \mu}, \quad e^2\kappa = \frac{\pi^2}3 k_B^2 T \sigma,
\ee
where $\sigma$ is the electrical conductivity. Taking $\sigma$ to vary linearly with carrier density, we find $Q$ in the disordered limit takes on the same qualitative form as Eq.(\ref{eq:cleanq}) in the clean limit. As a result, the qualitative features of energy-driven drag in the clean limit also appear in the disordered limit: namely, the avoided crossing of nodal lines (Fig. \ref{fig3} (a)), a peak at zero doping, double sign reversal along the diagonal $n_1=n_2$ and a three-peak structure along the diagonal $n_1=-n_2$ (Fig. \ref{fig3} (b,c)).
%, and the non-monotonic temperature dependence similar to the one displayed in Fig. \ref{fig3} (d). 
The $T$ dependence of $\rho_{21}$ is qualitatively similar in the diffusive and ballistic regimes. However, since the Wiedemann-Franz relation gives $\kappa\propto T$ (as opposed to $\kappa\propto 1/T$ in the ballistic regime), we find $\rho_{21}\propto T^2$ at lowest $T$ and $\rho_{21}\propto T^{-4}$ at higher $T>\mu,\Delta$,
as shown in Fig. \ref{fig3} (d).

In summary, vertical energy transfer in graphene heterostructures has strong impact on lateral charge transport in the Coulomb drag regime, dominating the drag response at low doping. Drag measurements thus afford a unique probe of energy transfer at the nanoscale, a fundamental process which is not easily amenable to more conventional techniques such as calorimetry, and is key for the physics of strong interactions that occur near neutrality.
% in the strong coupling regime is prominent because leakage of energy to the lattice is small. This regime is not easily probed by other methods such as calorimetry. Hence, energy-driven drag provides the means with which to study interlayer scattering and energy transfer which is particularly adept at revealing the physics of strong interaction that occur near neutrality.

%Energy-driven drag is a unique feature of Graphene Heterostructures and slowly varying density fluctuations caused by electron-hole puddles. Its signature is particularly striking close to DCN where a non-vanishing, positive drag with a DCN peak, and double sign reversals in drag make it easy to identify. This opens up new ways in which strong interlayer electron-electron interactions at DCN can be directly probed in transport. 

We acknowledge useful discussions with A. K. Geim, M. Serbyn and financial support from the NSS program, Singapore (JS) and the Office of Naval Research Grant No. N00014-09-1-0724 (LL).

\section{Appendix A: Heat current and Onsager reciprocity}

Here we use Onsager reciprocity to obtain thermopower induced by a temperature imbalance in a Fermi gas. This can be done in a general form applicable in both the ballistic regime, when the mean free path is dominated by electron-electron scattering, and in the diffusive regime, when the mean free path is dominated by elastic scattering by disorder. 

We start with recalling Onsager reciprocity for the heat and charge transport. Given charge current $\vec j$ and heat current $\vec j_q$ described by 
\bea\label{eq:onsager}
&& - \vec j = L_{11} \frac{1}{T} \nabla \mu + L_{12} \nabla \frac{1}{T}  
\\
&& \vec j_q = L_{21} \frac{1}{T} \nabla \mu + L_{22} \nabla \frac{1}{T}
,
\eea
the cross-couplings obey the Onsager relation $L_{12} = L_{21}$. Next, we consider the heat current, which can be obtained from the heat transport equation
%\addLL{Is specific heat missing?}
\be
\label{eq:heat}
\partial_t (C_{\rm el} T ) - \nabla \kappa \nabla T =  -\vec j \cdot \nabla F[n(\vec r),T ]
,
\ee
where $C_{\rm el}$ is the electron heat capacity, $F[n(\vec r),T ]$ is a function of the carrier density $n(\vec r)$ to be determined later, $T$ is the temperature and $\kappa$ is the thermal conductivity. From Eq.(\ref{eq:heat}) we find that the heat current is
\be\nonumber
\vec j_{\rm q} = - \kappa \nabla T +  \vec j F[n(\vec r),T ] = - \kappa \nabla T -  \sigma F[n(\vec r),T ] \nabla \mu
,
\ee
which gives $L_{21} = -T \sigma F[n(\vec r) ,T]$ in Eq.(\ref{eq:onsager}). %\addJS{ there seems to be a sign conflict with eq. \ref{eq:jq} above?}
Here we have used $\nabla \cdot \vec j = 0$ and $\vec j = - \sigma \nabla \mu$, where $\sigma  = L_{11}T$. Using the Onsager relation,
%(Eq. \ref{eq:onsager}), 
the thermopower e.m.f. induced by the temperature gradient equals
\be
\vec E = - \frac{F [ n(\vec r),T]}{T} \nabla T
.
\label{eq:thermopower}
\ee
This result has general validity irrespective of the transport mechanism specifics, which are manifested through the form of $F [n(\vec r),T]$. This quantity equals the Seebeck coefficient $s$ given by Eq.(\ref{eq:Mott_formula}) in the diffusive regime, whereas in the ballistic regime it is given by Eq.(\ref{eq:cleanq}).

%\addLL{Shall we provide derivation of Eq.(\ref{eq:cleanq})?} sure

The functional form of $F [n(\vec r),T]$ can be obtained by considering the kinetic equation (at steady state)
%
%To obtain the heat current as a function of applied bias or charge current we first consider the kinetic equation (at steady state)
\be
e\vec E \cdot \nabla_{\vec p} n(\vec p, \vec r) = I_{1} + I_{2}%, \quad \vec{j_q} = \sum (\epsilon-\mu) v_{\vec p} n(\vec p, \vec r), \quad \vec j = e \sum v_{\vec p} n(\vec p, \vec r)
\label{eq:kinetic}
\ee
where $e<0$ is the carrier charge. Here we write the collision integral as a sum of momentum non-conserving  and momentum conserving parts, $I_1+I_2$, corresponding to disorder scattering and electron-electron scattering, respectively.
%
% and $I_{i}$ are the collision integrals where $1$ denotes momentum non-conserving and $2$ denotes momentum conserving collisions.} 
% When $\vec E = 0$, $n(\vec p, \vec r) = n_0 = n_F (\epsilon - \mu)$ and no heat current or charge current flows. Here $n_F$ is the Fermi distribution. 
Heat and change current can be expressed through a steady-states deviation from the equilibrium Fermi distribution, $\delta n = n- n_0$, as follows
\be
\vec{j}_{\rm q} = \sum_{\vec p, i} (\epsilon_i-\mu) \vec{v}_{\vec p, i} \delta n_i(\vec p, \vec r), \quad \vec j = e \sum_{\vec p, i}\vec v_{\vec p, i} \delta n_i(\vec p, \vec r)
.
\label{eq:jqandj}
\ee
Here $i$ labels the conduction and valence band states and $\vec{v}_{\vec p, i}  = \p\epsilon_i/\p\vec p$ and $\epsilon_i = \pm v |\vec p|$. Below we consider the diffusive and ballistic regimes. In the first case the mean free path is dominated by elastic momentum non-conserving scattering  ($I_1$), in the second regime the mean free path is dominated by inelastic momentum-conserving scattering ($I_2$).
%two limits: (i) diffusive, and (ii) ballistic transport.

In the diffusive regime, neglecting $I_2$ and using the relaxation time approximation for $I_1$, we find
%\addJS{impurity scattering occurs far faster than electron-electron scattering so that we can neglect $I_2$ in Eq. \ref{eq:kinetic}. } Using the relaxation time approximation 
%
\be
\delta n_{\vec p} = e\vec{E} \cdot \vec v_{\vec p} \tau (\epsilon) \partial n_F / \partial \epsilon
,
\ee 
where $\tau$ describes elastic scattering by impurities.
This gives the standard expressions for Seebeck and Pelteir coefficient described in Eq.(\ref{eq:Mott_formula}) of the main text, so that $F = sT$.

%We briefly describe the derivation of Eq. \ref{eq:cleanq}.
In the ballistic regime, 
the fastest scattering mechanism comes from the (total) momentum conserving process of electron-electron scattering. As a result, we will neglect all other terms apart from $I_2$ in Eq. \ref{eq:kinetic} and look for distributions, $ n(\vec p, \vec r)$, that give a non-zero particle flow. At a nonzero total current, the non-equilibrium distribution can be written as $n_F(\epsilon-\vec p\cdot\vec u)$, where the term $\vec p\cdot\vec u$ describes the change due to particle flow. This allows us to write $\delta n$ as 
%Writing $\delta n = n - n_0$, where $n$ and $n_0$ are the non-equilibrium and equilibrium distributions respectively, we obtain the heat current
%evaluated by performing the Sommerfeld expansion:
%
\be
%\vec{j}_q = \sum_{\vec p}(\epsilon-\mu) v_{\vec p}\delta n_{\vec p} ,\quad 
 \delta n_{\vec{p} i} = -\vec u \cdot \vec p \frac{\partial n_F}{\partial \epsilon}  %=\frac{\vec u}{d}\int d\epsilon N(\epsilon) (\epsilon-\mu) vp\frac{\p n_F}{\p \epsilon}
\label{eq:deltan}
\ee
where the Fermi distribution, $n_F$, has a temperature that may depend on the flow and position. Using Eq. \ref{eq:deltan} and summing over both conduction and valence bands (where $\epsilon_c = v |\vec p|$ and $\epsilon_v = - v |\vec p|$, $c$ and $v$ refer to conduction and valence bands respectively) in Eq. \ref{eq:jqandj} we obtain
\bea
\vec j_{\rm q} &&= \frac{-\vec u }{2} \int_{- \infty}^{\infty} d\epsilon \nu(\epsilon) \epsilon (\epsilon -\mu) \frac{\partial n_F}{\partial \epsilon},  \nonumber \\
 \vec j &&=\frac{-\vec u e}{2} \int_{- \infty}^{\infty} d\epsilon \nu(\epsilon) \epsilon\frac{\partial n_F}{\partial \epsilon}
\eea
where $\nu(\epsilon) = 2 |\epsilon| / (\pi \hbar^2 v^2)$ is the total density of states and we take into account that $v_{\vec p}\cdot\vec p=\epsilon$ with $\epsilon$ positive (negative) for the conduction (valence) band. Since the function $\epsilon \nu(\epsilon) \propto \epsilon |\epsilon|$, the integral {\bf cannot} be evaluated for arbitrary ratio $k_BT/\mu$. Instead, we analyze the degenerate case, $\mu\gg k_BT$, in which case we %can approximate $\frac{\p n_F(\epsilon)}{\p\epsilon}=-\delta(\epsilon-\mu)$. We obtain
% with the help of the 
use the Sommerfeld expansion to obtain
\be
\vec j=\frac{e}{2} \mu \nu(\mu)\vec u, \quad \vec j_{\rm q} = 2\zeta(2)\nu(\mu)(k_BT)^2 \vec u
\ee
where we have used the identity $\int_0^\infty \frac{e^x x^2dx}{(e^x+1)^2}=\zeta(2)=\pi^2/6$. Comparing both expressions to eliminate $\vec u$ we obtain
\be
\vec{j}_{\rm q}=4\zeta(2)\frac{(k_BT)^2}{e\mu}\vec j
\ee 
%The nonanalytic dependence will be smeared for detuning values comparable to thermal broadening, $\mu\sim k_BT$.
The singularity at $\mu=0$ is smeared by broadening of the Dirac point due to disorder and thermal fluctuations. We can account for smearing via
\be
\vec{j}_{\rm q}=\frac{4\zeta(2)}{e}\frac{(k_BT)^2\mu}{\mu^2+\Delta^2(T)}\vec j
,\quad
\Delta^2(T)=\Delta_0^2+\eta (k_BT)^2
\ee
where $\Delta_0$ is the Dirac point width parameter. This yields $F [n(\vec r),T] = Q$, giving  Eq.(\ref{eq:cleanq}) of the main text.

\section{Appendix B: Interlayer Energy Relaxation}

In this Appendix, we derive the interlayer cooling power for the heterostructure system defined by the Hamiltonian in Eq.(\ref{eq:hamiltonian}). Using Fermi's golden rule, the scattering rate is
\be
W_{\vec{k_1}',\vec{k_1}}  = \frac{2\pi N}{\hbar} \mathop{ \sum_{\vec{k_2}, \vec{k_2}'}}_{\vec q} F_{\vec{k_2}, \vec{k_2}' } | V_\vec{q} |^2  f(\epsilon_\vec{k_2})[ 1- f(\epsilon_{\vec{k_2}'})] \delta_\epsilon \delta_1 \delta_2 
\ee
where $\delta_{\epsilon} = \delta \big(\epsilon_{\vec{k_1}', \vec{k_1}}  + \epsilon_{\vec{k_2}', \vec{k_2}}  \big)$, and $\delta_1 = \delta_{\vec{k_1}', \vec{k_1} + \vec q} $ and $\delta_2= \delta_{\vec{k_2}', \vec{k_2} - \vec q}$. Here $N=4$ is the number of spin/valley flavors, $\{1,2\}$ denote the different layers, $\epsilon_{\vec{k}', \vec{k}} =  \epsilon_{\vec k'} - \epsilon_{\vec k}$, and $ F_{\vec k, \vec k'}=|\la \vec k'\alpha|\vec k\beta\ra|^2$ is the coherence factor ($\alpha$, $\beta$ label states in the electron and hole Dirac cones). $V_\vec{q}$ is the screened inter-layer Coulomb interaction described below. The energy-loss power is
\be
\mathcal{J}= -N\sum_{\vec{k_1}, \vec{k_1'}} W_{\vec{k_1}', \vec{k_1}} (\epsilon_\vec{k_1'} - \epsilon_\vec{k_1}) f(\epsilon_\vec{k_1}) \big[ 1- f(\epsilon_\vec{k_1'})\big]  F_{\vec{k_1}, \vec{k_1}' } 
\label{cool}
\ee
We can simplify the evaluation of these sums by writing
$\delta \big(\epsilon_{\vec{k_1}', \vec{k_1}}  + \epsilon_{\vec{k_2}', \vec{k_2}}  \big) = \int_{-\infty}^\infty d\omega \delta \big(\epsilon_{\vec{k_1}', \vec{k_1}} - \omega\big) \delta \big(\omega+ \epsilon_{\vec{k_2}', \vec{k_2}}  \big) $
and using the identity
$
 f_s(\epsilon_\vec{k}) \big[ 1- f_s(\epsilon_\vec{k'})\big] =  \big(  f_s(\epsilon_\vec{k}) - f_s(\epsilon_\vec{k'})\big) \times \big( N_s(\epsilon_{\vec{k'},\vec{k}}) + 1\big)
$
where $N(\omega) = 1/ (e^{\omega/k_BT} - 1)$ is the Bose function taken at the electron temperature (of that particular layer) and $s = \{1,2\}$ denotes the layers. 
Using the quantities
\be
\chi_s''(\vec q, \omega) = N\sum_{\vec k} F_{\vec{k}, \vec{k} + \vec{q} } \big(  f_s(\epsilon_\vec{k}) - f_s(\epsilon_{\vec{k}+\vec q})\big) \delta \big(\epsilon_{\vec{k} + \vec{q}} - \epsilon_{\vec{k}} - \omega\big). 
\label{eq:chi}
\ee
where $\chi_s''(\vec q, \omega)  = \frac1{\pi}\Im\Pi(\vec q,\omega)$ is the imaginary part of the susceptibility and $\Pi(\vec q,\omega)=N\sum_{\vec k} F_{\vec{k}, \vec{k} + \vec{q} } \frac{ f_s(\epsilon_\vec{k}) - f_s(\epsilon_{\vec{k}+\vec q})}{\epsilon_{\vec{k} + \vec{q}} - \epsilon_{\vec{k}} - \omega-i0}$ is the polarization operator. Using these we can re-write the energy loss power as
\be\label{eq:J}
\mathcal{J} = \frac{\pi}{\hbar}\int d\omega   \omega  \big(N_2(\omega) - N_1(\omega)\big) \sum_{\vec q} | V_\vec{q} |^2 \chi_1''(\vec q, \omega) \chi_2''(\vec q, \omega)   
\ee
where we have noted that $N(-\omega) = -(1+ N(\omega))$ and used the fact that $\chi_s''(\vec q, \omega) = -\chi_s''(-\vec q, -\omega)$ so as to only keep the odd part of the product $N_2(-\omega) (N_1(\omega)+1)$. 

The energy transfer between the two layers is dependent on the Coulomb interaction $V_\vec{q}$ between the layers. We treat $V_\vec{q}$ by accounting for polarization in both layers and screening in the RPA approximation. The RPA-screened coulomb interaction is
\be\label{eq:RPA}
V_\vec{q} = \frac{V^0_{\vec q}}{1- V^0_{\vec q}( \Pi_1 (\vec q,\omega)+ \Pi_2(\vec{q,\omega}))}
,\quad 
V^0_\vec{q} = \frac{2\pi e^2 }{\varepsilon q}
\ee
where $\varepsilon$ is the background dielectric constant. Here we ignored the correction due to finite interlayer spacing $d$, approximating the interlayer interaction $V_\vec{q} = \frac{2\pi e^2}{\varepsilon q}e^{-d|q| }\approx V^0_\vec{q}$. This approximation is valid when the layer separation $d$ is small compared to the screening length in the layers, $d q_0\ll 1$, $q_0=-\Pi_1 (\vec q=0)- \Pi_2(\vec{q}=0)$. We will be interested in the regime when $d$ is small compared to the Fermi wavelength in the layers, for which the above approximation is adequate.

%The exact form of $\Pi(\vec q,\omega)$ depends on doping ($\mu\gg k_BT$ vs. $\mu\ll k_BT$) and on dynamical effects  [dynamical screening vs. static screening]. 
In the degenerate limit, $\mu\gg k_BT$, the polarization is given by 
\be\label{eq:PIdegenerate}
\Pi(q,\omega)%=-\nu\oint \frac{d\theta}{2\pi}\frac{qv\cos\theta}{qv\cos\theta-\omega-i0}
=-\nu(\mu) \lp 1-\frac{\omega}{\sqrt{(\omega+i0)^2-q^2v^2}}\rp
\ee
%In the static limit, $\omega\ll vq$, Eq.(\ref{eq:PIdegenerate}) yields the density of states, $\Pi=-\nu = -dN/d\mu$.
In the limit of $q \ll q_0$ where $q_0=(2\pi e^2/\varepsilon)(\nu_1+\nu_2)$, where $\nu_{1,2}$ is the density of states at the Fermi level in each layer, we can write $|V_q| = 1/| \Pi_1 (\vec q,\omega)+ \Pi_2(\vec{q,\omega})|$. Noting that in the degenerate limit, intra-band transitions are the dominant processes we approximate $\chi''(\vec q,\omega)$ by the imaginary part of Eq.(\ref{eq:PIdegenerate}). This allows us to write the cooling power as 
\be
\mathcal{J}_{12} = \frac{\pi}{\hbar}\int d\omega   \omega  \big(N_2(\omega) - N_1(\omega)\big) \sum_{|{\vec q} |>\omega/v} \frac{\nu_1\nu_2 \omega^2}{ (\nu_1 + \nu_2)^2 q^2v^2}
\ee
%= \frac{12 \zeta(4)}{2\hbar^3v^2} \frac{\nu_1\nu_2 k_B^4}{ |\nu_1 + \nu_2|^2}\Big( T_1^4 {\rm ln} \frac{T_0}{T_1}  - T_2^4\frac{T_0}{T_2}  \Big) 
%
%\frac{q^4}{(4\hbar)^2\pi^2} \frac{(2\pi e^2/\kappa)^2}{q^2(\omega^2-v^2q^2)+(4\pi e^2 q^2/4\hbar^2\kappa)^2} 
%\ee
where $T_0 = vq_0/k_B$. Using the idenity $\int_{-\infty}^{\infty} \omega^3 d\omega N(\omega) = 2 \cdot 3! \zeta(4) (k_BT)^4$, and recalling that the ultra-violet cutoff arises from screening via $q \ll q_0$ we obtain Eq.(\ref{eq:coolingpower}).

\end{document}